\documentclass[aps,prb,twocolumn,superscriptaddress,showpacs]{revtex4-1}
\usepackage{graphicx}
\usepackage{epstopdf}
\usepackage{color}
\definecolor{red}{rgb}{0,0,0}
\newcommand{\red}{\color{red}} 

\begin{document}

\title{EPR study of nano-structured graphite}

\author{Jonas Kausteklis}
\affiliation{Faculty of Physics, Vilnius University, Sauletekio al. 9, 10222 Vilnius, Lithuania}

\author{Pavel Cevc}
\affiliation{Institute "Jo\v{z}ef Stefan", Jamova cesta 39, 1000 Ljubljana, Slovenia}

\author{Denis Ar\v{c}on}
\email{denis.arcon@ijs.si}
\affiliation{Institute "Jo\v{z}ef Stefan", Jamova cesta 39, 1000 Ljubljana, Slovenia}
\affiliation{Faculty of Mathematics and Physics, University of Ljubljana, Jadranska 19, 1000 Ljubljana, Slovenia}

\author{Lucia Nasi}
\affiliation{IMEM-CNR, Parco Area delle Scienze 37/a, 43124 Parma, Italy}

\author{Daniele Pontiroli}
\affiliation{Physics Department, University of Parma, Via G. Usberti 7/a, 43124 Parma, Italy}

\author{Marcello Mazzani}
\affiliation{Physics Department, University of Parma, Via G. Usberti 7/a, 43124 Parma, Italy}

\author{Mauro Ricc\`{o}}
\affiliation{Physics Department, University of Parma, Via G. Usberti 7/a, 43124 Parma, Italy}

\date{\today}

\begin{abstract}
We report on a systematic temperature dependent X-band electron paramagnetic resonance (EPR) study of nano-sized graphite particles prepared by ball-milling.  In as-prepared  samples a very intense and sharp EPR resonance at $g=2.0035$ has been measured. The EPR linewidth shows a Korringa-like linear temperature dependence  arising due to the coexistence and  strong exchange coupling of itinerant and localized edge states. With a prolonged aging in inert atmosphere changes in the EPR signal suggest gradual structural reconstruction where the localized edge-states dominate the EPR signal. In this case the EPR spin susceptibility shows a maximum at $\approx 23\, {\rm K}$ indicating the development of antiferromagnetic correlations as expected for the graphene lattice with a bipartite symmetry.
\end{abstract}

\maketitle

% $$$$$$$$$$$$$$$$$$$$$$$$$$$$$$$$$$$$$$$

\section{Introduction}
 
The recent discovery of the fascinating electronic properties of graphene \cite{Novoselov04} and its possible application in the development of new electronic and spintronic devices initiated the study on how these properties are modified when the average dimensions of the $sp^2$ carbon system are reduced to the nano-scale. Consequently, the study of nano-graphene,\cite{Ritter} nano-sized multilayers \cite{Craciun} and nano-sized graphite has recently received a major boost. In particular nano-sized graphite, thanks to its possible large-scale production, large specific area and reactivity, can meet applications as gas sorbents, in particular for hydrogen storage,\cite{Orimo} high temperature gaskets and even as electrode for lithium recharged batteries.\cite{Wang}

In all nano-sized graphitic particles the structure {is rather defective with influences on their physical and chemical properties.\cite{Dress} Apart from the obvious increase of edges, also ripples, vacancies, Stone Wales or similar  are created during the sample preparation resulting} in a plethora of intriguing localized electronic states  that co-exist with the itinerant electrons. Most surprisingly, high-temperature carbon-based magnetism is repeatedly reported\cite{Makarova}   and associated with such defect states. For instance, dangling bonds or vacant atoms were suggested to induce magnetism in amorphous carbon samples,\cite{aCH1, aCH2} including carbon nanofoam.\cite{Cfoam1, Cfoam2} Ferromagnetism has been also found after irradiation of highly oriented pyrolitic graphite by protons of energy 2.25 MeV.\cite{Pablo}
In graphene layers, zig-zag edge states appear remarkably stable, are ferromagnetically polarized and coupled to itinerant states\cite{Wakaba} and are considered as a prime suspect for the observed magnetic hysteresis loop in many graphene samples.\cite{FMgrap1, FMgrapn} 

As it is well known, the reduction of the graphite grain size down to the few nanometers range, even if performed through mechanical processing like milling, does not bring to its amorphization but nanocrystalline phases are obtained.\cite{Welham} This is confirmed by TEM and SEM analysis of the ball-milled graphite which show the formation of nano-sized graphite platelets in which the $A-B$ stacking order of graphite is probably lost although preserving the pristine layered structure. On the other hand, at moderate milling time, no relevant damage of the honeycomb intra-layer order is observed.\cite{disorder} 
It is  well known that reactivity of dangling bonds at graphite edges indicates that both the nano-sizing process and the successive manipulation need to be performed under strict oxygen/moisture free conditions. However, even in such atmosphere, the edge states may still be unstable and thus inclined towards the edge reconstruction as it is  currently discussed in the literature both on a theoretical as well as on experimental level.\cite{Radovic, Koskinen}
 
Our understanding of the physical and chemical properties of nano-sized graphite platelets thus depends on the answers to two important open questions:  (i) What is the origin of edge states, their interactions with the itinerant electrons and possible consequences for the carbon-based magnetism? and (ii) What would be the edge reconstruction that affects their long term stability and how is it reflected in the magnetic response? We therefore decided to prepare nano-sized graphite samples by a ball-milling process and then employ electron paramagnetic resonance (EPR)  since this technique  proved to be extremely sensitive local-probe technique for the study of $\pi-$electron states {\red in different forms of carbon.\cite{Barklie}} We found that the edge states  couple to itinerant electrons giving rise to exchange-narrowed EPR spectra. Indications for their antiferromagnetic rather than ferromagnetic ordering have been found at low temperatures. The concentration of these states is immediately suppressed by at least an order of magnitude when samples were exposed to air speaking for their high reactivity. Surprisingly, we found that the EPR signal also changes on a time scale of several months even if the samples are kept in inert atmosphere. This implies very slow edge reconstruction processes  towards the  configuration where the antiferromagnetic exchange becomes even stronger. Possible stable edge configurations and consequences for the magnetism in nano-sized graphite are discussed. 

% $$$$$$$$$$$$$$$$$$$$$$$$$$$$$$$$$$$$$$$

\section{Methods}

Nano-sized graphite samples were prepared starting from the graphite powder of high purity, RW-A grade, initial mesh 240, from SGL Carbon. About 500 mg of graphite powder, which was treated at 830 $^\circ$C overnight in dynamical vacuum, were placed in a ZrO$_2$ grinding bowl (bowl volume 15 ml) and ball-milled in Ar atmosphere ($<1$ ppm O$_2$ and H$_2$O) for 1 h at 50 Hz with a Fritsch Mini-mill pulverisette 23. Milling process was stopped every 10 minutes in order to avoid bowl overheating. No ZrO$_2$ contamination was observed in any of the prepared samples.  Extreme care was taken that ball-milled graphite (BMG) samples were not exposed to air at any stage of preparation or handling.

{\red 
Special care was taken to avoid the formation of amorphous carbon as some degree of amorphization can also be obtained after prolonged mechanical milling. It is well known that nano-crystalline graphite is obtained after 8 h milling time, while if exceeding this time some fraction of carbon starts to turn to amorphous state.\cite{disorder} For this reason we ball-milled graphite powder for 1 h, i.e. a time significantly shorter from that required for amorphization. Raman spectroscopy is in principle  one of the most appropriate techniques for discriminating between nano-crystalline graphite and amorphous carbon.\cite{Ferr}  An extensive Raman investigation of ball-milled samples show, that the G peak position is around 1580 cm$^{-1}$,\cite{MakarovaR} which is a typical value for graphite and quite far from the value of 1510 cm$^{-1}$ expected for amorphous carbon.\cite{Ferr}
}

\begin{figure} [tbp]
\includegraphics[width=0.93\linewidth, trim=120 20 120 20, clip=true]{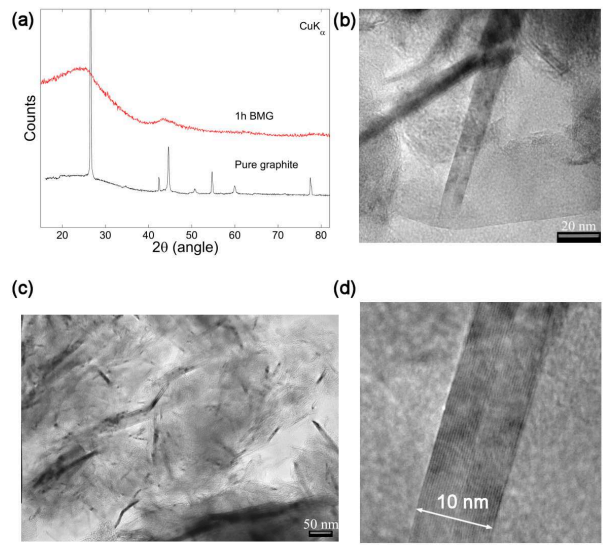}
\caption{(color online). (a) Comparison between the powder X-ray diffractograms of the starting pure graphite (bottom in black) and the 1 h ball-milled samples (top in red). (b) TEM images of the ball milled graphite sample shows that the sample consists of larger graphite nano-sized platelets and smaller graphite nanoribbons. (c) Nanoribbons, whose mean dimensions along the stacking $c$-axis is 6-7 nm, appear crystalline and (d) preserve the initial graphite stacking distance of $\sim$0.35 nm. }
\label{XRD}
\end{figure}

Nano-sized graphite samples were also checked by the laboratory XRD analysis, collected at room temperature on a Bruker D8 Discover diffractometer with GADDS (Cu$K_\alpha$). The ball-milling causes the loss of correlation among graphene planes (turbostratic disorder), since the majority of the reflections disappeared, except for (0 0 $l$) reflections, which are  strongly broadened (Fig. \ref{XRD}a). In particular, (0 0 2) peak appear slightly downshifted, very broadened and asymmetrical, as expected in turbostratic graphite.\cite{Warren} The TEM images obtained with a Jeol 2000FX microscope working at 200 kV show that high energy ball-milling on graphite promotes  the  formation of graphite nano-sized platelets as well as nanoribbons (Fig. \ref{XRD}b,c) {in agreement with previously reported BMG characterizations.\cite{Welham}} The mean size along the stacking $c$-axis is around 6-7 nm (mean over 50 observations). The stacking distance of the planes inside a ribbon was found to be $\sim$0.35 nm (see Fig. \ref{XRD}d), which is in very good agreement with the value obtained from XRD data on the starting graphite powder ($d_{stacking}$ = 0.3557(4) nm). This suggests that the milling process does not introduce significant in-plane defects (delaminations, changes of curvature of the planes), which should increase the interlayer distance, as observed in other similar studies.\cite{Welham}  

Continuous wave (cw) EPR experiments were performed on the home built X-band EPR spectrometer equipped with Varian TEM104 dual cavity and Oxford cryogenics ESR900 cryostat. Experiments were carried in the temperature range between 300 and 5 K with a temperature stability better than $\pm 0.1\, {\rm K}$. Additional measurements were also performed at different microwave powers ranging between 0.1 and 5 mW  in order to exclude the effect of  saturation on the EPR signal.

% $$$$$$$$$$$$$$$$$$$$$$$$$$$$$$$$$$$$$$$

\section{Results and Discussion} 

\begin{figure} [tbp]
\includegraphics[width=0.85\linewidth, trim=0 0 0 0, clip=true]{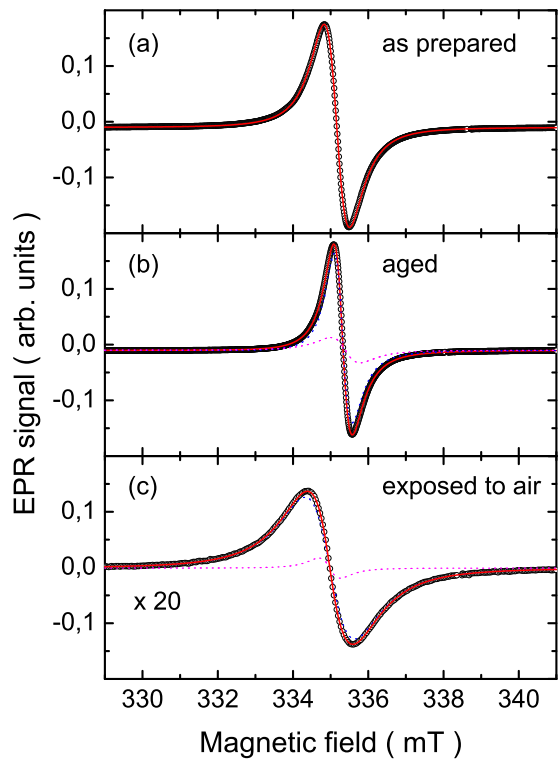}
\caption{(color online). Comparison of X-band EPR spectra measured on ball-milled graphite samples collected at different stages of sample-treatments, i.e. (a) as-prepared, (b) aged in an Ar atmosphere for several months, and (c) immediately after exposure to air. The spectra were collected at 100 K. In all three cases open circles represent the experimental data, solid red line is a fit, while dotted lines represent the individual components used to fit the spectra.}
\label{EPRspectra}
\end{figure}

In Fig. \ref{EPRspectra} we show EPR spectra ($T=100\, {\rm K}$) of samples prepared with a ball-milling process but treated in a different way. In as-prepared sample we find a very intense and extremely narrow line that can be well simulated with a single Lorentzian component. The measured $g-$factor of 2.0035(2) is very close to the electron free value and thus confirms that this resonance is due to the carbon-based states. Magnetic field scanning over broader field-range did not reveal any additional resonances that would belong to frequently observed transition-metal impurities. The half-width full-maximum $\Delta H_{1/2}$ of this resonance is only 0.11 mT and it is much less than it is usually reported for graphene-like samples.\cite{LCiric,Forro} In fact it suggests strong exchange narrowing limit in this case, as we shall demonstrate later on. 

The EPR spectra drastically change if the samples are subjected to  different treatments. As one can immediately notice from Fig. \ref{EPRspectra}b,   if the sample is aged for several months in inert atmosphere, the EPR spectrum  slightly narrows and becomes asymmetric because the additional resonance starts to appear. We stress that the intensity of the EPR signal does not significantly decrease with time implying that  the paramagnetic centers do not die with time, but rather slowly redistribute. The main component is now even narrower than in the as-prepared sample and  $\Delta H_{1/2}=0.08\, {\rm mT}$ is measured. The  narrowing of the main EPR component suggests  stronger exchange-narrowing limit. The  minor component is rather similar to the one measured in the as-prepared sample with the linewidth of 0.12 mT and is thus likely a residual of the original signal.    

Changes are severely  more dramatic  when the as-prepared sample is exposed to air. Immediately after the exposure the EPR signal intensity decreases by nearly an order of magnitude. This clearly proves that the vast majority of paramagnetic centers, which we observed in as-prepared sample, is highly reactive and promptly react with the oxygen- or hydrogen including functional groups to create C--H, C--OH or similar EPR silent groups.  Therefore, the remnant EPR component is now  broader, i.e. $\Delta H_{1/2}=0.21\, {\rm mT}$, while the narrow one with $\Delta H_{1/2}=0.08\, {\rm mT}$ has almost disappeared (Fig. \ref{EPRspectra}c). 

\begin{figure} [t]
\includegraphics[width=0.955\linewidth, trim=0 0 0 0, clip=true]{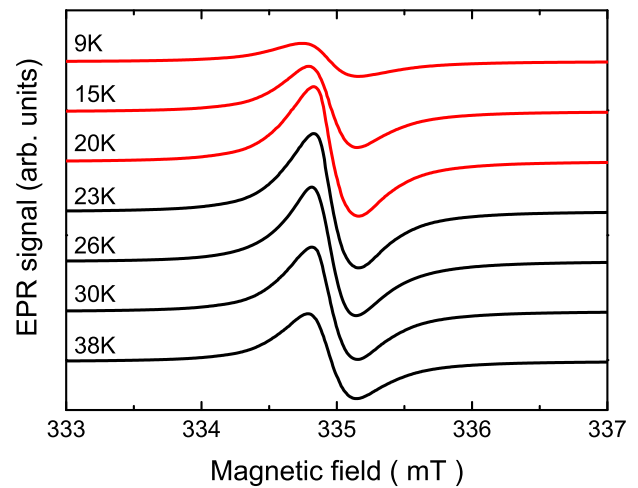}
\caption{(color online). Temperature dependence of the X-band EPR spectra in ball-milled graphite aged in inert atmosphere for a prolonged time. Please note gradual broadening and decrease in intensity of the spectra below $\sim 23$ K.}
\label{Tdep_spectra}
\end{figure}

The measured EPR spectra are  strongly temperature dependent. In order to demonstrate this we show in Fig. \ref{Tdep_spectra} the low-temperature dependence measured in aged ball-milled graphite. In this temperature range EPR spectra first get narrower on cooling but below $\approx 23\, {\rm K}$ they start to broaden again. At the same time the spectral intensity starts to decrease and the spectrum measured at 9 K is already by a factor of 3 weaker than the one taken at 23 K.  Since the EPR spectra are so narrow, we also checked for the possible microwave saturation effects. We thus measured the temperature dependence of EPR spectra at different microwave powers (from 0.1 to 5 mW) and observed the same kind of behavior in all cases. These additional experiments unambiguously prove that the disappearance and broadening of the EPR spectra at low temperatures is an intrinsic effect and most likely signals the development of local magnetic fields and antiferromagnetic correlations{\red \cite{Gatt}} in the ball milled graphite. Similar behavior has been observed also in as-prepared sample except that signal disappearance is shifted to much lower temperatures (not shown). On the other hand, the sample that was exposed to air completely lacks such behavior.

\begin{figure} [tbp]
\includegraphics[width=1\linewidth, trim=0 -5 0 -20, clip=true]{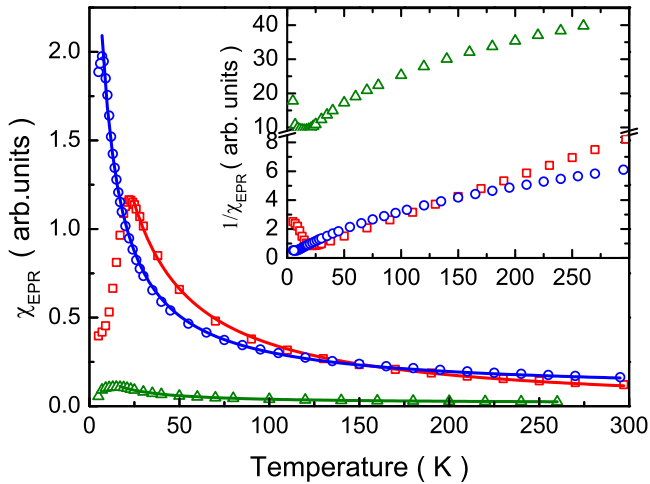}
\caption{(color online). Temperature dependences of the  EPR spin susceptibilities for the as-prepared (circles), aged in inert atmosphere (squares) and air exposed (triangles). Solid lines represent fits to a sum of Pauli temperature independent and Curie-Weiss spin susceptibilities (see text for details). Inset: Temperature dependences of the  inverse EPR spin susceptibilities  for the same samples. }
\label{chi_EPR}
\end{figure}

Since the EPR spectra indicate some sort of magnetic ordering at low temperatures we now turn to the intensity of the EPR signal, $\chi_{EPR}$, which is directly proportional to the spin-only susceptibility. In Fig. \ref{chi_EPR} we summarize its temperature dependence for all three samples, i.e. as-prepared, aged and air-exposed. Although $\chi_{EPR}$  increases with decreasing temperature for as-prepared sample it cannot be  described by the Curie-Weiss, $\chi_{CW}$,  dependence only and additional temperature independent Pauli-like susceptibility, $\chi_P$, has to be taken into account as well. The $\chi_{EPR}$ data has been thus simulated as a sum of the two contributions, i.e.  
$
\chi_{EPR} (T)=\chi_{P}+C/(T-\theta)\, .
$
Since it was impossible to reliably measure the mass of the investigated EPR samples, we could not quantitatively determine $\chi_P$ nor the Curie constant $C$. However, the relative ratio between the Pauli and Curie-Weiss components can be determined and at room temperature it amounts $\chi_{CW}/\chi_P\approx 0.7$ implying that the contribution of itinerant electrons is substantial. The negative Curie-Weiss temperature $\theta=-2.6(5)\, {\rm K}$ indicates weak antiferromagnetic interactions between magnetic moments. The sharp maximum in $\chi_{EPR}$ seen at 9 K (Fig. \ref{chi_EPR}) is thus signaling antiferromagnetic correlations  below this temperature.  When the samples are exposed to air, both contributions to $\chi_{EPR}$ are affected. Not surprisingly, the indications for the antiferromagnetic correlations also disappear  meaning that the magnetic ordering is essentially associated with the carbon states and not to some non-carbon impurities. 

Aging in inert atmosphere  suppresses $\chi_P$ and now $\chi_{CW}$ is the dominant component of the spin susceptibility. Simultaneously, the antiferromagnetic exchange interactions become stronger as  the Curie-Weiss temperature increases to $\theta=-6(1)\, {\rm K}$. The temperature, where $\chi_{EPR}$ shows a maximum, increases to $\approx 23\, {\rm K}$. The higher temperature where antiferromagnetic correlations develop, corroborates well with enhanced $\theta$.   The comparison of $\chi_{EPR}$ measured for as-prepared and aged sample  suggest that the localized edge states in as-prepared samples are structurally and electronically unstable and that the edges in nano-sized graphitic particles  slowly reconstruct on a time-scale of several months. The final structure seems to converge towards the structure where the exchange interactions between the localized edge-states become stronger.  

\begin{figure} [tbp]
\includegraphics[width=1\linewidth, trim=0 0 0 0, clip=true]{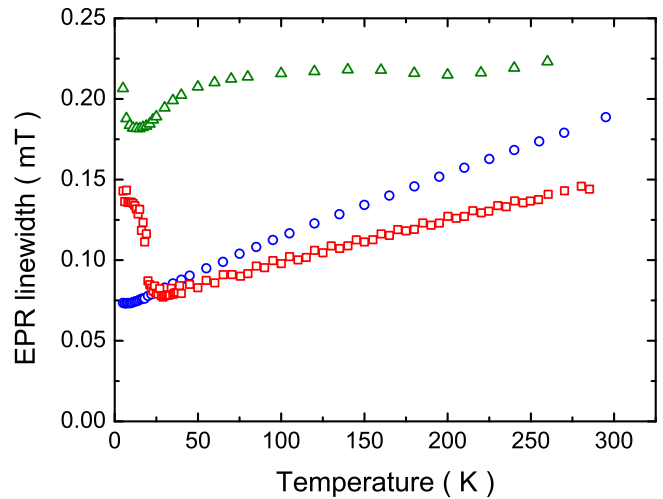}
\caption{(color online). Temperature dependences of the  EPR linewidth for the as-prepared (circles), aged in inert atmosphere (squares) and air exposed (triangles). }
\label{dH}
\end{figure}

The above discussion  proved that nano-sized graphite samples contain both itinerant as well as localized edge-states and that at low temperatures antiferromagnetic correlations develop between localized moments. Such situation should greatly affect the spin dynamics and should be thus reflected in the temperature dependence of the EPR linewidth, $\Delta H_{1/2}$.  The temperature dependence of $\Delta H_{1/2}$ is shown for all three samples in Fig. \ref{dH}. It is immediately evident that $\Delta H_{1/2}$ has completely different behavior in air-exposed samples compared to the other two samples. Both, for as-prepared as well as for aged samples the linewidth shows a pronounced linear temperature dependence. Since we argued earlier from the temperature dependence of the EPR spin susceptibility that localized and itinerant electrons coexist we take this dependence of   $\Delta H_{1/2}$ as an indication for the Korringa-type of relaxation, i.e. $\Delta H_{1/2}={\pi k_B\over g\mu_B}\left[ JN(E_F)\right]^2 T=aT$. Here $J$ is the exchange coupling constant  and $N(E_F)$ is the density of states at the Fermi level. The extracted Korringa constants $a=3.9\times 10^{-4}\, {\rm mT/K}$ and $a=2.8\times 10^{-4}\, {\rm mT/K}$ for the as-prepared and aged samples respectively  are understandably very small due to the small values of $N(E_F)$. The antiferromagnetic ordering is seen as a sudden broadening of the resonance. This is particularly well expressed in the aged sample, where $\Delta H_{1/2}$ below $T_N=23\, {\rm K}$ follows a  dependence that resembles a temperature dependence of the order parameter. Therefore,  the moments, that give rise to the residual  EPR signal below $T_N$ are sensitive to  the growing local magnetic fields in the ordered state. However, we notice that we were not able to detect the (anti)ferromagnetic resonance modes probably due to considerable disorder leading to their large linewidth.  The temperature dependence of $\Delta H_{1/2}$ for the air-exposed sample on the other hand does not indicate that Korringa relation would dominate the spin dynamics. In fact, the observed  $\Delta H_{1/2}$ resembles weak temperature dependence of the EPR linewidth found in some disordered carbon nanoparticles with  negligibly  antiferromagnetic correlations at low temperatures.

The present results on ball-milled graphite demonstrate that the localized states with $g=2.0035$ are responsible for the magnetism in these samples. We next notice, that  $g-$factors in the range between 2.0030 and 2.0045 have been  frequently measured in polymer-like\cite{poly1, poly2} or diamond-like films\cite{diam} in environments where a mixture of $sp^2$ and  $sp^3$-hybridized states has been suggested. One of the possibilities for such re-hybridization in  nano-sized graphitic particles would be for instance the binding of the interstitial carbon atoms, similarly as in irradiated graphite\cite{Yazyev} but clearly more work is needed to clarify this point. 
Intriguingly, the as-prepared nano-sized graphitic particles seem to be electronically and structurally unstable and some slow edge reconstruction must take place. Such structural reconstruction may be also associated with  the changes from the $A-A$ stacking {or turbostratic} to the $A-B$ stacking. This hypothesis is further supported by our recent Raman investigation  
on the same materials where the evolution of the 2D peak shape with  sample aging has been correlated with the readjustment of the  graphite stacking.\cite{MakarovaR} 
The main result of this work is the finding of antiferromagnetic-like reduction of the EPR spin susceptibility at low temperatures  in a highly disordered nano-sized graphitic particles where complete lack of spatial correlations between defects is expected. Since the graphene lattice has a bipartite symmetry and  antiferromagnetic correlations between the two sublattices have been predicted by several theoretical studies,\cite{Yazyev, Yazyev2} then the total magnetization arising from the two sublattice-resolved average magnetic moments should vanish as the defects equally distribute over both sublattices.\cite{Yazyev}  

% $$$$$$$$$$$$$$$$$$$$$$$$$$$$$$$$$$$$$$$

\section{Conclusions}

In conclusion, a systematic EPR study of nano-sized graphite samples prepared by ball-milling process was conducted.  In as-prepared samples localized edge-states coexist and couple to itinerant electrons leading to extremely narrow EPR resonances. With a prolonged aging even in inert atmosphere changes in the EPR signal  suggest  gradual edge reconstruction that may also  involve changes in the stacking of graphene layers. In parallel with structural evolution a (partial) localization of itinerant electrons and  stronger antiferromagnetic interactions lead to the development of antiferromagnetic correlations below 23 K. Our results thus fully comply with the  bipartite nature of the graphene layers where vanishing total magnetization is expected when defects equally distribute over both sublattices.   The majority of the EPR active edge-sites created during the sample preparation were found to be very air-sensitive suggesting high chemical reactivity of such edge states. 
%The proposed scenario of magnetic ordering together with the ferromagnetism in zig-zag edge states in graphene demonstrates that the carbon-only magnetism is a reality rather than a dream.   

\begin{acknowledgments}

M.R., D.P. and M.M. acknowledge the financial support from the Swiss National Science Foundation HyCarBo project (grant No. CRSII2-130509).

\end{acknowledgments}

\end{document}